\documentclass{article}
\usepackage{amssymb}
\usepackage{amsbsy}
\usepackage{epsfig}

\begin{document}

\noindent The Roger Tayler Memorial Lectures  1999-2003 {\it Astron. Geophys.} (special issue, 2005) pp 16--25

\begin{center}
\huge{Towards a theory of rapidly oscillating Ap stars}

\vspace{0.5cm}


\large{Douglas Gough,\\ Institute of Astronomy, Madingley Road,
  Cambridge, CB3 0HA \& Department of Applied Mathematics and
  Theoretical Physics, Centre for Mathematical Sciences, Wilberforce
  Road, Cambridge, CB3 0WA.}

\vspace{1cm}
\normalsize
{Printed: 28 March 2005}

\end{center}

\begin{abstract}
Peculiar A stars are so named because they exhibit abundance
peculiarities in their atmospheres.  It is believed that these arise
as a result of differentiation of chemical species in large magnetic
spots in which convective mixing is inhibited: there might be just two
antipodal spots, whose axis is inclined to the axis of rotation.  Many
of the Ap stars that are rotating slowly also pulsate, with periods
substantially shorter than the period of the fundamental radial mode.
The pulsations appear to be nonradial, but axisymmetric, with their
common axis usually aligned with the axis of the spots.  In this
lecture I shall first discuss the magnetic suppression of convection
in the spots, and then I shall try to explain the pulsation phenomenon, 
reviewing some of the suggestions that have been made to explain the 
alignment and the excitation mechanism, and finally raising some issues 
that need to be addressed. 
\end{abstract}

\section{Introduction}

It was my privilege to be the first research student to be registered
by the University of Cambridge under Roger Tayler's supervision. That
did not make me Roger's first student in practice, however, because
Roger was a very kindly man and would supervise other astrophysics
students in the department who needed more help than their registered
supervisors provided; and so there were already several students under
Roger's wing when I arrived. Consequently Roger was at that time
intellectually active in a variety of fields, embracing stellar
structure and evolution, the physics of thermonuclear reactions,
magnetohydrodynamics -- particularly stability theory, with which
Roger had a great deal of experience -- and stellar pulsation. It is
therefore appropriate that I devote this lecture to a subject that
combines many of these fields. I have chosen to address some of the
issues that arise from trying to understand the phenomenon of rapid
oscillations in peculiar A stars. Roger did not himself work
 in this area, but, recognizing his interests, one can easily
imagine that he might well have done so had his attention not been
drawn more strongly to other matters.

My own interest in rapidly oscillating Ap (roAp) stars was triggered
by Donald Lynden-Bell, when in 1981, if I recall correctly, he told me
on his return from a visit to the South African Astronomical
Observatory of Don Kurtz's recent, and yet unpublished,
discovery. What intrigued me was that the stars were spotted and that
the oscillations were apparently dipolar, with axes aligned with the
spots. There were only a few examples, but nevertheless they must
surely have at least hinted very strongly that the oscillations are always (almost) aligned with the
spots, and therefore do not wander far off under the influence of
Coriolis precession. Soon after that time, N\"{o}el Dolez visited me
from Paris, and we decided to carry out some
simple calculations that might address the most obvious questions: why
are only dipolar oscillations excited; why are their axes aligned with
the spots; and why are only rapid oscillations observed, with
frequencies much higher than that of the fundamental dipole mode? I
address unashamedly in this lecture the picture that is emerging at
least in my mind from that early work, and which has been
substantially elaborated upon in recent years; I cannot here also review in
detail the alternative suggestions that have been propounded. And I
emphasize here that the line of argument that I follow is strongly influenced by
what I learnt at Roger's feet when I was a student.

\section{The roAp-star phenomenon}

Rapidly oscillating Ap stars are relatively slowly rotating A stars
with chemical abundance peculiarities: they are rich particularly in
Sr, Cr and Eu. The stars probably lie somewhat above the main sequence
in the HR diagram, more-or-less in the pulsational instability
strip. Their masses and radii are both roughly twice solar. They have
convective cores, which will not concern me here, and shallow
convection zones immediately beneath their surfaces. They appear to
pulsate in high-overtone nonradial p modes of low degree, with periods
generally in the range 4--15 minutes, the period of fundamental radial
oscillation being about two hours. Typically only a very few modes are
detected, sometimes only one, although HR 1217 is richly endowed with
at least seven.  The oscillation amplitudes, in
radiative intensity, are typically of the order of 1mmag. This is some
thousand times greater than the greatest amplitudes of stochastically
excited intrinsically stable acoustic oscillations in stars like the
Sun, so one might presume that the modes are intrinsically
overstable. In most of the roAp stars that have been observed there
are just one or two frequencies present in the spectrum of the
oscillations (together, perhaps, with one or two higher harmonics); in
a few others there are several. Fine structure, attributed to degeneracy splitting modulated by rotation, is usually present. The oscillations are often apparently
dipole modes, although sometimes quadrupolar components appear to be present
too. Many Ap stars have large-scale, probably mainly dipolar, magnetic
fields with field intensities ranging from several hundred Gauss to a few 
kiloGauss. The axis of the field is not, in general, aligned with
the axis of rotation. For more details the reader is referred to the
excellent reviews by Kurtz(1990, 1995) and Cunha (1998).

\begin{figure} \label{fig1}
\centering
\resizebox{0.65\textwidth}{!}{\includegraphics{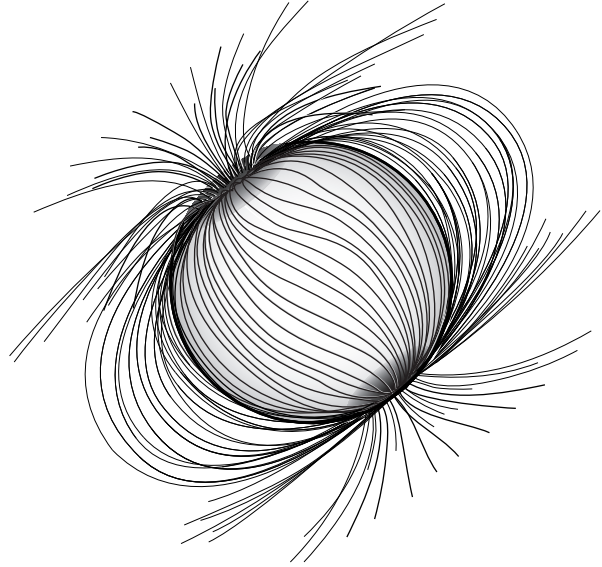}}
\caption{A rapidly oscillating Ap star, with some lines of its
  external magnetic field.}
\end{figure} 

The picture of an Ap star that one might have in mind is perhaps
somewhat like Figure 1, which illustrates the star and some of
the lines of its external magnetic field. This field matches onto an
essentially dipole field pervading the interior of the star
(except, perhaps, the convective core). Near the magnetic axis, where
the field is nearly vertical, convection is presumed to be suppressed
by the field, probably throughout the entire radial extent of the
unstably stratified zone. This permits chemical element segregation, by a
combination of radiative levitation and gravitational settling against
diffusion and by advection by a putative stellar wind (e.g. Vauclair,
Dolez and Gough, 1991), leading to the appearance of two large antipodal spots
of chemical peculiarity each of which might occupy some 10 or 20 per
cent of its hemisphere. Severe line blanketing inhibits the radiant heat
output, redistributing the optical spectrum in such a manner as to
reduce the heat flux in the frequency ranges in which the stars are
normally observed. However, the total heat flux is not altered
substantially: because the spots are large, and because lateral heat
transfer in the radiative region beneath is ineffectual, the radial
stratification must adjust to let the heat out, unlike in sunspots.  The large-scale magnetic field pervading
the stellar interior presumably causes the star to rotate uniformly;
Lorentz stresses induced by putative differential rotation would act
to oppose the shear on a timescale of only a few years. 

Because the star is rotating, the spots can, in
principle, be mapped by Doppler imaging; two recent examples of 
Doppler-imaged surface chemical-element abundance variations are presented in
Figure 2.

\begin{figure} \label{fig2}
\centering
\resizebox{0.75\textwidth}{!}{\includegraphics{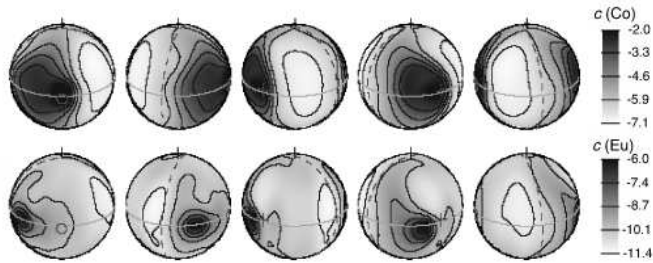}}
\caption{Distribution of the concentration $c$ of cobalt (upper row) and
  europium (lower row) over the surface of the roAp star HR 3831,
  inferred via Doppler imaging by Kochukhov {\it et al}. (2004).
Shown are the star in five uniformly spaced rotation phases; a meridian
is indicated by the dashed curve to indicate the relative orientation   
of the star.  Contours are plotted at intervals of 1.0 dex; the
darker areas are the regions of high concentration.  The distribution
is predominantly dipolar, with axis almost in the (rotational)
equatorial plane (which is indicated in the diagrams).  Matching a
dipole to the magnetic-field data yield a polar surface field strength
of 2.5kG inclined from the axis of rotation by 87$^{\rm o}$.  The axis of
rotation is inclined from the line of sight by 68$^{\rm o}$.}
\end{figure}

As I mentioned in the introduction, a startling property of the
oscillations is that they appear to be axisymmetric, with axes
coinciding with the location of the spots. This kinematical
description was first proferred by Kurtz (1982), who called it the
{\textit{oblique pulsator}} model.  But there must be a dynamical
explanation; surely, there must be an interaction with the magnetic
field, either direct or indirect, which accounts for the alignment.  A
direct interaction might cause one of the three independent dipole
modes of any given order to be aligned with the field; if that were
so, there would remain the problem of explaining why that and only
that member of the triplet is excited to an observable amplitude.  The
most likely explanation is that conditions in the spot, where the
amplitude of the aligned mode is greatest, are conducive to mode
excitation, whereas conditions elsewhere are not.  Before addressing
that central issue, I first offer some justification for the
presumption that the magnetic field actually succeeds in suppressing
convection in the spots.

\section{The influence of a magnetic field on convection}

This is the subject on which I worked with Roger when I was a
student. Our goal was to find, by linearized perturbation analysis,
conditions for the stabilization by a simple magnetic field of an
otherwise convectively unstably stratified region of a star.  In the
event we approximated the potentially unstable region by a
plane-parallel layer of ideal gas in hydrostatic equilibrium, and, for
tractability, we chose the field to be either uniform but
arbitrarily inclined, or vertical but varying with a single horizontal
Cartesian coordinate \textit{x}. Chandrasekhar (1961) had written extensively
on a corresponding problem for a Boussinesq liquid, but there were
few results for a compressible gas -- results only for the case when
the field is horizontal.

I had already learned from Roger's lectures that energy arguments can
be powerful in establishing sufficient conditions for stability, so I
set about establishing an appropriate energy integral for the problem
in hand. I think I must have worked for the better part of the eight
weeks of a Cambridge Full Term \footnote{`Full Term' in Cambridge is
  the period within Term during which university lecturers, like Roger
  was at the time, were almost fully distracted from research by
  teaching and administration.}
without discussing the matter seriously
with Roger, but by the end I had succeeded in establishing an integral
relation, which, together with an immediate deduction from it, I
proudly presented to Roger when Full Term was over. I had had some
very brief discussions with Roger over coffee during those weeks, and
had learned that Roger had made some progress himself, so I was eager
to discover whether Roger too had derived such an integral, and if so,
how far he had progressed in extracting from it some useful
criteria. It turned out, to my dismay, that he was well ahead of me. But when we compared
notes, I was horrified to discover that Roger had not even attempted
to construct an energy integral like the one that I had so
painstakingly derived, but instead had simply written it down at the
start of his investigation. He had done so simply by specializing a
more general result that had been published in a well known (by plasma
physicists) paper by Bernstein, Frieman, Kruskal and Kulsrud (1958),
but of which I was quite unaware. Why had Roger not told me about
it? Because, it subsequently transpired, Roger knew that I would
appreciate the result more by deriving it myself than I would by
merely reading it. Moreover, it taught me, as I realized only later,
an important lesson about teaching, which I have since used to the benefit
of my own students -- although I'm not sure that at the time I implement  
it my students necessarily describe the service as a benefit!

\subsection{A physical introduction}

Before proceeding with the discussion, permit me first to anticipate
the result. As is evident from the discussion in the paper that Roger
and I wrote about this investigation (Gough and Tayler, 1966) and a
subsequent discussion concerned explicitly with roAp stars (Balmforth
{\it et al}., 2001), the ability of a field to inhibit convection
depends not only on its intensity but also on its orientation. A rough
idea of the condition required of a magnetic field to stabilize an
otherwise convectively unstable fluid layer can be obtained by
comparing estimates of the (destabilizing) buoyancy force associated
with a perturbation with the (stabilizing) restoring Lorentz
force. Evidently, what one must demand immediately is that the
magnetic field configuration be itself intrinsically stable; we are
not interested here in the possible stabilization by a (presumably
convectively stable) fluid layer of an otherwise unstable magnetic
field: that is a subject on which Roger worked, with others,
later. The buoyancy force per unit mass tending to drive convection is
roughly minus the product of the square of the (imaginary) 
buoyancy frequency {\it{N}} and
the vertical displacement $\delta r$ of the fluid, namely $-N^2\delta
r = (\Gamma^{-1}-\gamma^{-1})g^{2} \rho/p$.  Here, $g$ is the local
acceleration due to gravity, $p$ is pressure and $\rho$ is density;
$\Gamma = {\rm{d \,ln}}p / {\rm{d \,ln}} \rho$ is a measure of the
stratification of the background state, and $\gamma = (\partial \,
{\rm{lnp}} /\partial {\rm{ln}} \rho)_{s}$, the partial derivative being taken at constant
specific entropy $s$, is the first adiabatic exponent.  This
overestimates the destabilizing force somewhat, because it
ignores the opposing pressure gradients associated with the horizontal
component of the displacement of the fluid, which must necessarily be
present. The magnetic restoring force per unit mass exerted on the
displaced fluid, in SI units, is roughly $(B^2k^2/\mu\rho)\delta r$,
where $B$ is the field strength, $k$ is a characteristic wavenumber of the
magnetic field-line distortion, and
$\mu$ is the magnetic permeability of the fluid. Thus one
anticipates that for the field to prevent convection $v^2\gtrsim -k^{-2}N^2~$,
where $v=B/\sqrt{\mu\rho}$ is the Alfv\'en speed. The criterion
therefore depends critically on the greatest permissible lengthscale,
$k^{-1}_{\rm m}$, of the perturbation.  That depends on the field geometry.

At the magnetic poles the field is almost vertical, and the
characteristic wavenumber of the field-line perturbation is associated
with the horizontal component of the putative motion. The scale of
that motion is determined partly by an appropriate scale height of the
background state, which for simplicity I take to be the pressure scale
height $H_p = p/g \rho$,
and partly by the depth $d$ of the unstably stratified zone: one might
expect $k^{-1}_{\rm m}={\rm min}(H_p, d/2\pi)$. In a typical roAp
star, $H_p<d/2\pi$, and consequently it is required that $v^2 \gtrsim
-H^2_p N^2$ for stability. This condition may be re-written 

\begin{equation} \label{1}
{1 \over \Gamma}-{1 \over \gamma} \lesssim {B^2 \over \mu p}\ . 
\end{equation}

\noindent
It is satisfied in the model illustrated in Figure 3, for example, for
$B$ of order $10^{-1}$T (1 kG), which is not atypical of roAp-star
fields. Thus it seems not unreasonable that convection is suppressed
near the magnetic poles.  In the equatorial regions, on the other hand, 
if the magnetic field is
predominantly dipolar, not only is the field somewhat weaker (by a factor of 2) than it is near the poles, but
also it is nearly horizontal, and is therefore distorted by the
vertical component of the convective motion, which is limited by
neither the vertical scale heights nor the depth of the convection
zone. Indeed, in principle the length-scale could be very much larger than either, as had been found by
Newcomb (1961) and Tayler (1961). In fact,
Newcomb and Tayler found that a purely horizontal field in a
horizontally infinite fluid layer did not influence the criterion for
convective instability at all; and indeed sufficient conditions for stability
derived subsequently by Roger and me (Gough and Tayler, 1966) depend
only on the vertical component of the field. It is therefore much less
likely that convection is suppressed near the magnetic equator;
Lorentz stresses might succeed in imposing some degree of horizontal
coherence to the motion, and in the nonlinear regime may even reduce
the heat-flux-to-temperature-gradient ratio, thereby raising the
temperature gradient above that in a corresponding nonmagnetic star. 
But it is likely that the unstably stratified layers are fully
convective near the magnetic equator.

\subsection{Locally applicable sufficient conditions for stability}

Having painted the picture, let us now add a little rigour, and
address a formal stability problem: a problem similar to those
addressed by Roger and me, but tailored to suit the situation in
hand. I adopt a simple spot model: namely, a plane layer of fluid,
infinite in horizontal extent, and in hydrostatic equilibrium under
gravity, pervaded by a vertical magnetic field that is axisymmetric
about a vertical axis. For the purposes of discussing any putative
instability, which I presume to occur on a dynamical timescale, the
fluid is taken to have no viscosity and vanishing magnetic and
thermal diffusion coefficients, although it must be acknowledged that
diffusion was necessary for setting up the equilibrium state in the
first place. The work
$\delta W$ required to effect a displacement
$\boldsymbol{\xi}(r,\phi,z)=(\xi_r,\xi_{\phi},\xi_z)$, with respect to
cylindrical polar coordinates with $z$ vertical (and increasing
downwards), is most easily derived from the work of Bernstein,
Frieman, Kruskal and Kulsrud (1958), and may be written:

\parbox{10cm}
{\begin{eqnarray*} 
\delta W
 = \frac{1}{2}\int \left[ \right. g\xi_z\nabla\rho \cdot \boldsymbol{\xi} & + &
  2g\rho\xi_z{\rm div}\boldsymbol{\xi}+(\gamma p+B^2)({\rm
    div}\boldsymbol{\xi})^2 \\
& - &  2B^2 \xi^\prime_z{\rm div}\boldsymbol{\xi}+B^2
\boldsymbol{\xi}^{\prime} \cdot
\boldsymbol{\xi}^{\prime}\left. \right] {\rm d}V \nonumber
\end{eqnarray*}}
\hfill
\parbox{1cm}
{\begin{eqnarray}\label{2}
\end{eqnarray}}

\begin{equation}   \label{3}
\equiv \delta \tilde{W}+\frac{1}{2}\int B^2\xi^{\prime 2}_\phi {\rm d}V~,
\end{equation}
in which $B(r)$ is the equilibrium magnetic field (now in units in
which the magnetic permeability is unity), $p$ and $\rho$ are the
equilibrium pressure and density, and a prime denotes differentiation with respect to $z$.  
The integration is over the volume of the fluid, which I presume
to be a convectively unstable (in the absence of a magnetic field)
layer sandwiched between two stable layers.  It is obvious on
energetic grounds that the perturbation must decay with distance from
the convectively unstable layer;  therefore, I choose two boundaries,
at $z=z_1$ and $z=z_2$ well into the stable layers, and set
$\boldsymbol{\xi}=\boldsymbol{0}$ on these boundaries.  Also, I should
point out that the convectively unstable layer need not be
convectively unstable throughout; in the specific model considered
below, it comprises two unstable zones separated by a stable zone.

If $\delta W > 0$ for all admissible functions $\boldsymbol{\xi}$ (e.g., all suitably differentiable vector functions that vanish on
the boundaries), then no displacement of the fluid is possible without
the exertion of work by external forces, and the equilibrium state
must be intrinsically stable.  On the other hand, if there exist
admissible functions $\boldsymbol{\xi}$ for which $\delta W < 0$, one
cannot infer instability, because those displacements may not be
realizable: only displacements that satisfy the equations of
magnetohydrodynamics (mhd) are actually permitted. Thus, from the energy
integral one can derive only sufficient conditions for stability. But
that will be adequate for the purpose here. 

To obtain conditions that
are both necessary and sufficient one must study the linearized mhd
equations. A powerful way to proceed, in general, is via a variational
principle derived from those equations, which can be used to bound
below the growth rates $\eta$ of the linear modes. Much use of the
method has been made by Chandrasekhar (1961). In the case considered here, the sign of $\eta$
rests on the sign of the integral (\ref{2}), so the energy method and the
variational method look very similar. It is still necessary to
restrict attention to permissible functions $\boldsymbol{\xi}$; and
often this can be accomplished most readily by considering the
eigenfunctions of the linearized equations, which satisfy the
Euler-Lagrange equations for the stationarity of $\eta$: if a
criterion can be found to render at least one eigenvalue $\eta$
positive, then the system is unstable. In most cases, to find a single
condition that is both necessary and sufficient requires solving the
Euler-Lagrange equations. This can be accomplished analytically in the
absence of a magnetic field, and leads to the local criterion that
Schwarzschild (1906) introduced into the astronomical literature
(Lebowitz, 1965, 1966; Gough, 2001). However,
no such criterion exists for the case being discussed here, because a
magnetic field connects one region of the fluid to another, and the
necessary and sufficient condition must necessarily be nonlocal, and 
possibly global. However, I shall
not pursue that matter any further here, because the analysis I do
present yields a sufficient condition for stability that is useful enough for
discussing Ap-star spots.

Before proceeding with the analysis it is instructive (or perhaps
merely only reassuring) to make a straightforward qualitative
observation: the coefficient of $B^2$ in the integrand in equation 
(\ref{2}) is 
positive definite. That implies that at least a state that is neutrally
stable when $B=0$ is rendered stable by the introduction of the
field.  Moreover, it is suggestive, although not yet
demonstrated, that a convectively unstable state can be stabilized by
a sufficiently intense field.

The objective of the analysis is thus to find a simple bound
on $B$ above which the minimum value of $\delta W$ amongst all displacements
$\boldsymbol{\xi}$ is positive.  Evidently that condition on $\delta
  W$ will be satisfied if $\delta
\tilde{W} > 0$. The displacement $\boldsymbol{\xi}$ can then be
considered to be characterized by the functions $\xi_r,~\xi_z$ and
div$\boldsymbol{\xi}$. The Euler-Lagrange equation for minimizing
$\delta \tilde{W}$ with respect to div$\boldsymbol{\xi}$ is 

\begin{equation}  \label{4}
\left( \gamma p + B^2 \right) {\rm div}\boldsymbol{\xi} = B^2\xi^\prime_z - g\rho\xi_z~,
\end{equation} 
which can then be substituted into the formula for $\delta
\tilde{W}$. It is evident that substantial simplification can be
achieved by restricting attention to spot models in which $\nabla\rho$
is vertical: then

\parbox{10cm}
{\begin{eqnarray*} 
\delta\tilde{W} & = & \frac{1}{2}\int \left[ g\rho^\prime\xi^2_z -
  \left( \gamma
p+B^2 \right) ^{-1}\left( B^2\xi^\prime_z-g\rho\xi_z \right) ^2+B^2 \left(
  \xi^{\prime2}_z+\xi^{\prime 2} _r \right) \right] {\rm d}V \\
& \equiv & \delta\hat{W} +\frac{1}{2}\int B^2\xi^2_r {\rm d}V~, \nonumber
\end{eqnarray*}}
\hfill
\parbox{1cm}
{\begin{eqnarray}\label{5}
\end{eqnarray}} 
which is positive if $\delta \hat{W} > 0$; the functional $\delta
\hat{W}$ depends on $\boldsymbol{\xi}$ only through $\xi_z$ and its
$z$ derivative (there are no horizontal derivatives), and therefore the condition $\delta\hat{W} > 0$ can be
applied separately on each field line, and, rather than being a
partial differential equation, the minimizing Euler-Lagrange equation
is then merely an ordinary differential equation. The condition for
$\delta \hat{W} > 0$ can therefore be written:

\begin{equation} \label{6}
{\cal{E}} \left( B^2; \xi_z \right)  \equiv \int^{z_2}_{z_1} \left( F\xi^2_z + G\xi^{\prime2}_z
\right) {\rm d} z >
0~~~{\rm on~every~field~line},
\end{equation}
where

\begin{equation}   \label{7}
F=-\left(\frac{\gamma g \rho p}{\gamma p +B^2} \right)^\prime +
\frac{g^2\rho^2}{\gamma p+B^2} 
+g^\prime\rho~,
\end{equation}

\begin{equation} \label{8}
G=\frac{\gamma pB^2}{\gamma p+B^2}~,
\end{equation}
subject, of course, to the hydrostatic constraint ${\rm d}p/{\rm d}z =g\rho$. An
immediately obvious condition is $F>0$ everywhere. However, it is more
useful to introduce a function $\Phi$, which at present is arbitrary,
and rewrite inequality (6) as

\begin{equation} \label{9}
{\cal{E}} = \int^{z_2}_{z_1}\left[ \left( F-\Phi^\prime
-G^{-1}\Phi^2 \right) \xi^2_z+G \left( \xi^\prime_z-G^{-1}\Phi\xi_z
\right)^2\right]{\rm d}z ~,
\end{equation}


\noindent for then a more general stability criterion can be written
down, namely

\begin{equation} \label{10}
F-\Phi^\prime-G^{-1}\Phi^2>0 \qquad{\rm everywhere.}
\end{equation}
In particular, because $\gamma$ is a rapidly varying function of depth
in the ionization zones of H and He, one might choose to set

\begin{equation} \label{11}
\Phi = \frac{-g\rho B^2}{\gamma p+B^2}+\Psi
\end{equation}
in which $\Psi$ does not depend explicitly on derivatives of
$\gamma$ or $g$. In that case, condition (\ref{10}) does not depend
explicitly on $\gamma^\prime$ or $g^\prime$. Then, for example, one could set
$\Psi=g\rho B^2/(\alpha p+B^2)$, where $\alpha$ is constant, as did
Gough and Tayler (1966), to obtain, if $\alpha=1$,

\begin{equation}  \label{12}
\frac{1}{\gamma}-\frac{1}{\Gamma} < \frac{B^2}{\gamma \left( p+B^2
  \right) }~,
\end{equation}
or

\begin{equation} \label{13}
\frac{1}{\gamma}-\frac{1}{\Gamma}<\left(1+\frac{1}{\gamma}
-\frac{1}{\gamma_{\rm max}}\right) \frac{B^2}{\gamma_{\rm max}p+B^2}~,
\end{equation}
if $\alpha=\gamma_{\rm max}$, $\gamma_{\rm max}$ being the greatest
value of $\gamma$ in the region.  Both criteria have a superficial
resemblance to the approximate condition (\ref{1}). The system is
stable if either of these strict conditions is satisfied
everywhere. It should be remarked that the left-hand sides of the
conditions can be written in terms of the
dimensionless superadiabatic temperature gradient
$\nabla-\nabla_{\rm ad}$, which is more familiar to those working in the
theory of stellar structure. For a gas composed principally of
hydrogen and helium, the required relation is approximately
$\gamma^{-1}-\Gamma^{-1} =  -(\nabla-\nabla_{\rm ad}-\hat{\mu}{\rm
d}\ln\mu_0/{\rm d}\ln p) \hat{\delta}$, where $\hat{\delta}= -(\partial \ln \rho
/\partial \ln T)_{p},_{\mu_0}>0$ is a dimensionless isobaric thermal expansion
coefficient and $\hat{\mu}=(\partial\ln T/\partial\ln
\mu_0)_{p,\rho}$, $\mu_0$ being the mean molecular mass of the fluid
when it is completely
unionized. For a perfect gas, $\hat{\delta}=1$ and $\hat{\mu}=1$.

Although conditions (\ref{12}) and (\ref{13}) can be applied locally,
they are actually global conditions, because stability is assured only
if either of them is satisfied everywhere. However, some influence of
the nonlocality has been thrown away by ignoring the contribution from
$\xi_z^{\prime 2}$ to $\cal{E}$, in addition to having already ignored
the contribution from $\xi_\phi$ and part of the contribution from
$\xi_r$ to $\delta W$. In Figure 3 is plotted
$\gamma^{-1}-\Gamma^{-1}$, together with the right-hand sides of the
inequalities (\ref{12}) and (\ref{13}) with $B = 1 {\rm kG}$; they are 
plotted  through the
outer layers of an Ap-star envelope model in radiative equilibrium
similar to the model with the small accumulation parameter used by Balmforth {\it et al.,} (2001) to represent a spot.  Although the right-hand
sides of both criteria (\ref{12}) and (\ref{13}) exceed the left-hand
sides nearly everywhere, they fail to do so in the hydrogen ionization
zone, in which there is a sharp density inversion. Moreover,
augmenting the magnetic field does not change matters significantly,
because both right-hand sides are bounded above by values that do not
differ substantially from unity. Consequently, stability against
convection is not assured by those locally applicable criteria. More
account must be taken of the nonlocal interactions. For example, it is
hardly likely that the term in the integrand in equation (\ref{9})
that was jettisoned in deriving criteria (\ref{12}) and (\ref{13})
vanishes in the interval in which the criteria are not satisfied, and
it is even likely to contribute substantially to stabilization,
particularly when $B$ is large. But to demonstrate that requires more
detailed analysis.

\subsection{A global sufficient condition}

Progress can be made in a relatively simple manner by investigating
more carefully the conditions under which the stability condition
(\ref{6}) is satisfied. To this end I seek the minimum of $\cal{E}$
amongst all admissible functions $\xi_z(z)$, and then find the
smallest value of $B$ for which that minimum is positive. This is
equivalent to demanding that the minimum value of 

\begin{equation} \label{14}
\lambda(B^2)=\frac{\cal{E}}{\int^{z_2}_{z_1} G\xi^2_z{\rm d}z}
\end{equation}
be positive, for which the Euler-Lagrange equation is

\begin{equation} \label{15}
\frac{\rm d}{{\rm d}z}\left(G\frac{{\rm d}\xi_z}{{\rm
d}z}\right)- \left( F-\lambda G \right) \xi_z =0~.
\end{equation}
Equation (\ref{15}) is to be solved subject to the boundary conditions
$\xi_z=0$ at $z=z_1$, and $z=z_2$. Solving this eigenvalue problem is much
simpler than solving the full stability problem, which, even for the
idealized configuration considered here (an axisymmetric vertical
magnetic field) is represented by a sixth-order partial
differential equation in two dimensions ($z$ and $r$).

\begin{figure} \label{fig3}
\centering
\resizebox{0.85\textwidth}{!}{\includegraphics{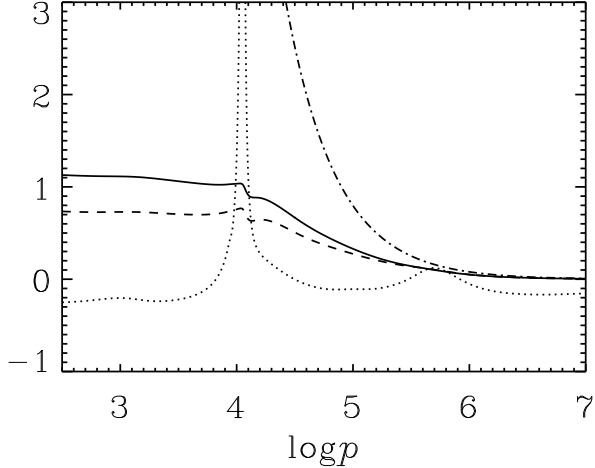}}
\caption{Convective stability characteristics of the polar region of
  an Ap-star model similar to the model with low accumulation
  parameter considered by Balmforth {\it et al}. (2001).  The dotted
  curve is $\gamma^{-1} - \Gamma^{-1}$, and the dot-dashed curve is
  the right-hand side of the approximate criterion (1) for a vertical
  field of strength ${\it B}=$1kG.  The solid and dashed curves are the
  right-hand sides of the sufficient conditions (12) and (13),
  respectively, also with ${\it B}$= 1kG.  The hydrogen and first
  helium ionization zones (10\%--90\% ionization), which are merged, lie
  in the interval $\log\!{\it p}=(3.7,4.9)$, the second helium
  ionization zone in $\log\!{\it p} = (5.3, 7.0)$.}
\end{figure}

I have solved equation (\ref{15}) for the spot model illustrated in Figure 3,
taking $z_1=0$ and $z_2$ to be many scale heights beneath the lower
region in which $\gamma^{-1}-\Gamma^{-1}>0$ (i.e. well below the
region where the stratification would be convectively unstable in the
absence of a magnetic field). The result is that $\lambda(B^2)>0$ for
all $|B|>B_{\rm c} =1.11{\rm kG}$; this critical magnetic field was
found to be essentially independent of $z_1$ and $z_2$ provided that the
boundaries are far enough away from the unstably stratified
regions. Because it is only a sufficient condition, this result
implies that the region is linearly stable to convection for all
vertical magnetic fields of strength greater than $B_{\rm
c}$. Thus, it is probably safe to conclude that convection is
suppressed in regions of Ap stars that are pervaded by kiloGauss
magnetic fields. Some Ap stars have been observed to have fields
several times greater than that.

\subsection{On the magnetic inhibition of convection}

It is important to realize that this calculation does not prove that
convection cannot occur in a spot. First, the analysis is linear, and
it has been shown that under some circumstances 
instability can occur as a subcritical direct bifurcation; that is to
say, convection at finite (i.e. not infinitesimal) amplitude can be
sustained against more intense mean fields than the critical linearly
stabilizing field. The mechanism is one of sweeping aside the field
into columns between convective cells, leaving convection to run its
course in the regions of relatively weak field between. It has been
seen to occur in numerical simulations of convection in a Boussinesq
fluid (Blanchflower and Weiss, 2002), and I see no reason why it 
should not occur in
compressible convection too. Second, the field strength declines away
from the centre of the spot, and there must be a radius beyond which
convection can take place. Although formally the analysis requires the
sufficient condition $\lambda (B^2)>0$ to be satisfied everywhere,
the fact that the criterion is applied separately on each field line
suggests that lateral interactions have been adequately accounted for
by the jettisoning of the term in $\delta W$ proportional to
$\hat{\xi}^{\prime 2}_\phi$ and by adopting the consequent Euler-Lagrange
equation (\ref{4}) relating the horizontal and vertical components of
the displacement, and that convection is suppressed (at least
linearly) where $\lambda >0$, and not necessarily elsewhere. Indeed,
our knowledge of the structure of sunspots adds some credence to this
idea. Moreover, Ap-star spots are laterally very extensive compared to
their depth, and it is hard to imagine that conditions at the lateral
boundary have a material influence on convection dynamics in the
middle. It is worth pointing out that the reduction of the sufficient
conditions for stability to criteria on separate field lines suggests
that the original assumption of axisymmetry is unimportant, and that the 
criteria are probably valid for any horizontal variation of $B$. This is
supported by the observation that the same criteria apply to vertical
fields that depend on a single horizontal Cartesian coordinate, as was
shown by Gough and Tayler (1966).

In conclusion, it is very likely that convection is strongly
inhibited, if not suppressed entirely, in the magnetic polar regions
of Ap stars with kiloGauss fields, permitting the creation of large
spots in which chemical segregation can lead to photospheric abundance
anomalies (Vauclair, Dolez and Gough, 1991). That necessarily leads to a
horizontal variation of the stratification, which has an influence on
the structure of the
eigenmodes of oscillation in addition to the direct influence by the
field itself via the Lorentz stresses.
It lifts the degeneracy amongst modes of like order and degree and so
determines the orientations of the modes relative to the magnetic and
the rotation axes, as will be discussed in \S6.  Is that at
least part of the reason why it appears that only
modes aligned with the magnetic poles are observed? The other part of
the reason
would then be that it is the spots that lead to mode excitation.  It is 
to these matters that I now turn my attention.

\section{Mode excitation in starspots}
I have already presumed that the modes of oscillation are likely to be
overstable, driven principally in the convectively stable spots.  The
most likely source of excitation is the $\kappa$ mechanism in
ionization zones, as is considered to be the case in the classical
variables; outside the spots convection dominates the
heat transport and thereby
quenches the $\kappa$ mechanism.  The key issue is to what extent
convection drives or damps the oscillations outside the spots, via the pulsation-induced
modulation of the convective heat-flux and the out-of-phase component
of the Reynolds stresses.  The issue was investigated by Dolez and
Gough (1982) and Balmforth {\textit{et al}}., (2001), using a
composite Ap-star model comprising a (double) conical region of a
spherically symmetrical stellar model with convection suppressed to
represent the spot, or polar, regions, and a complementary, equatorial
region of a corresponding normal stellar model in which convection was
treated using a time-dependent mixing-length formalism.  The direct
influence of the perturbed Lorentz stresses on the oscillations was
ignored.  In both investigations it was found that monople modes are
damped in the equatorial region, and it was argued that because the
work integral characterizing the damping and driving varies significantly
in only the outer layers of the star, the same conclusion must hold
also for any other low-degree mode.

\begin{figure} \label{fig4}
\centering
\resizebox{0.48\textwidth}{!}{\includegraphics{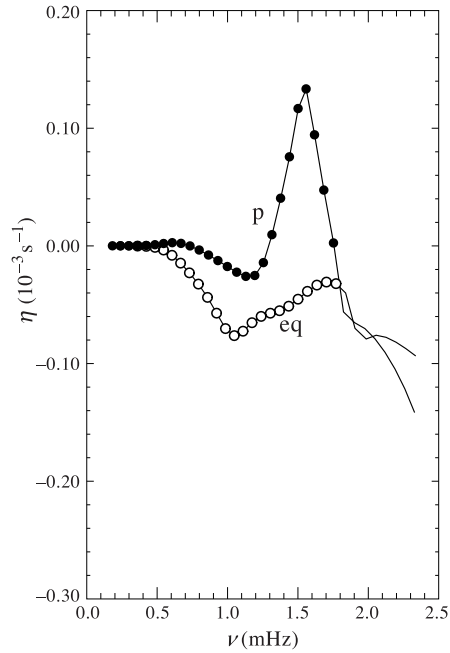}}
\caption{Contributions to the growth rates of radial modes from the
  polar and equatorial regions of the Ap-star model with $ M/{\rm
  M}_\odot = 1.87~, \log T_{e\!f\!f} = 3.910$ and $\log L/{\rm L}_\odot =
  1.164$ considered by Balmforth {\it et al}. (2001).  The growth rate
  of a global low-degree mode of azimuthal order $m$ with respect to the
  spot axis is obtained as the weighted average $\Lambda^m_l
  \eta^{\rm{p}} + (1- \Lambda^m_l) \eta^{{\rm{eq}}}$, where
  $\Lambda^m_l$ is depicted in Figure 5.}
\end{figure}

Studying the spot regions is somewhat complicated, because it is
likely that a stellar wind emanates from the spots, outwardly
advecting chemical species which are otherwise segregated by a
combination of radiative levitation and gravitational settling against
diffusion.  It is not unlikely that the balance of these processes
leads to a concentration of the helium in the vicinity of the zone of
first ionization of helium (Vauclair, Dolez and Gough, 1991), which
influences both the frequencies and the stability of the acoustic
modes. In particular, it reduces, perhaps surprisingly, the excitation
of high-order modes; but it adds to the driving of low-order modes.
Despite the uncertainties in the model, it was concluded by Balmforth
{\textit{et al}}., that the spot regions are likely to contribute
substantially to the driving of some of the modes, particularly some
modes of relatively high frequency, as is observed. This was found to
be the case even when the boundary conditions permitted energy leakage
into the atmosphere, in contrast to the earlier findings by Gautschy
and Saio (1998). Growth rates $\eta$
for complete (i.e. spherical) polar and equatorial models are
illustrated in Figure 4.  Whether the mode is globally excited or
damped depends, therefore, on the relative contributions to the growth
rate from the polar and the equatorial regions.  Before embarking on a
discussion of that matter, I first warn that the growth rates plotted
in Figure 4 are likely to be overestimated, because the perturbed
Lorentz forces were omitted from the calculations; in the surface
layers where Lorentz forces are significant the acoustic motion
couples with Alfv\'{e}n waves which propagate downwards and
subsequently dissipate in the deep interior of the star, thereby
extracting energy from the modes (Roberts and Soward, 1983).  However,
the estimates of the energy loss by Bigot {\textit{et al}}., (2000),
Cunha and Gough (2000) and Saio and Gautschy (2004) are too low for this
process to overcome the driving computed by Balmforth {\textit{et
al}}.~ (2001); therefore, with sufficiently large spots, one might
expect some of the global modes to be overstable.

In order to estimate the growth rates of the global modes it is
necessary to combine the separate calculations for the polar and the
equatorial regions.  The frequency $\nu_{nml}$ and the growth rate
$\eta _{nml}$ of a mode of order $n$, degree $l$ and azimuthal order $m$ are given approximately by

\begin{equation} \label{16}
\sigma_{nml} = \int^{2 \pi}_0 \left\{\sigma _{nlm}^{\rm {p}}
  \int^{1}_{\tilde{\mu}} \left|S_{nlm} \right|^2 {\rm{d}} \mu +
  \sigma^{\rm{eq}}_{nlm} \int ^{\tilde{\mu}}_0 \left| S_{nlm} \right|^2
  {\rm{d}}\mu \right\} {\rm{d}} \phi~,
\end{equation}
with respect to spherical angles $(\theta, \phi)$ about the axis of
symmetry of the spots.  Here, ${\textstyle\mu = \cos \theta,~
  \sigma_{nlm}= 2\pi \nu_{nlm} + {\rm{i}}\eta_{nlm}}$ is the complex frequency of the mode,
the scalar components $\Psi_{nlm}$ of the eigenfunction of which are
presumed to be factorized in the form $\textstyle\Psi_{nlm}(r,\theta, \phi, t) =
\psi_{nlm}(r) S_{nlm}(\theta, \phi){\rm{e}}^{-{\rm{i}} \sigma_{nlm}t}$, in which
$S_{nlm}$ is normalized such that $\int\!\! \int \left|S_{nlm} \right|^2
{\rm{d}} \mu {\rm{d}} \phi = 1$, the integration being over the entire
sphere, and the superscripts $\mbox{\rm{p and eq}}$ denote corresponding
  values of $\sigma_{nlm}$ for spherically symmetrical stellar models
  that are entirely polar-like or entirely equator-like,
  respectively; in the composite Ap-star model, the spots occupy the
  regions $|\mu | > \tilde{\mu}$.  
Were the modes to have been
  adiabatic, and the boundary condition perfectly reflecting,
  $\sigma_{nlm}$ would have been real and the pulsation equations
  self-adjoint, and equation 
(\ref{16}) 
would follow immediately from  the associated variational principle, provided that the acoustic
  differences between the polar and equatorial parts of the
 model were not too great (\textit{cf}., Cunha and Gough, 2000). But the
 pulsations considered here are actually nonadiabatic and are only
 partially contained within the leaky surface of the star.
 Nevertheless, a variational principle can be constructed from a
 combination of the solutions of the actual problem and its adjoint,
 and once again equation (\ref{16}) follows. Of course, in either case
 the function $S_{nlm}(\theta, \phi)$ should describe the actual
 eigenfunction, but, in view of the variational property of equation
 (\ref{16}), it may be replaced by the unperturbed eigenfunctions
 $Y_{lm}(\theta'', \phi'') = P^m_l (\mu''){\rm{e}}^{im \phi''}, P_l^m$
 being the associated Legendre function of the first kind (with
 argument $\mu''= \cos \theta''$) with respect to polar angles
 $(\theta'', \phi'')$ about the principal axis of the pulsations.  The
 orientation of that principal axis will be discussed in \S6.

Suffice it to say now that if the frequency perturbation associated
with the spots is much greater than that due to rotation, the mode
axis is more-or-less aligned with the spots, in accord with Kurtz's
(1982) original assumption.  In that case

\begin{equation} \label{17}
\displaystyle\sigma_{nml} \simeq \Lambda^m_l ~\sigma^{\rm{p}}_{nl} + \left( 1- \Lambda^m_l
\right)~ \sigma^{\rm{eq}}_{nl}~,
\end{equation}
where

\begin{equation} \label{18}
\Lambda^m_l = \left(2l+1 \right) \frac{(l-m)!}{(l+m)!}
\int^{1}_{\tilde{\mu}} \left(P^m_l \right)^2 {\rm{d}}\mu~,
\end{equation}
and
$\sigma^p_{nl}$ and $\sigma^{\rm{eq}}_{nl}$ are respectively
the ($m$-degenerate) complex eigenfrequencies of the corresponding
modes of the entirely polar-like and the entirely equator-like models.
Since for high-order modes $\nu_{nl} \simeq \left(n+{1 \over 2}l
\right)\nu_0~,$ where $\nu_0 \simeq {1 \over 2} \left( \int c^{-1}{\rm{d}}r
\right ) ^{-1}$ in which $c$ is sound speed, and the excitation
  and damping take place predominantly in the outer layers of the star
  -- and, one must note, the relative deviation of the inertia of the
  mode from that of the corresponding radial mode is only
  ${\rm{O}}\! \left( l^2/n^2 \right)$ -- one can express the cyclic
  frequency $\nu_{nlm}$ and the growth rate $\eta_{nlm}$ in terms of the values for
  the corresponding radial modes:

\begin{eqnarray} \label{19}
\nu_{nlm} & \simeq & \left(1+l/2n \right) \left[ \nu^{\rm{eq}}_{n 0} +
  \left( \nu^{\rm{p}}_{n{\rm{0}}}- \nu^{\rm{eq}}_{n 0} \right)\Lambda^m_l
  \right]~,\\
& & \nonumber\\
\eta_{nlm} & \simeq & \Lambda^m_l ~\eta^{\rm{p}}_{n 0} + \left( 1-\Lambda^m_l
  \right)\eta^{\rm{eq}}_{n 0}~.
\end{eqnarray} \label{20}

\begin{figure} \label{fig5}
\centering
\resizebox{0.69\textwidth}{!}{\includegraphics{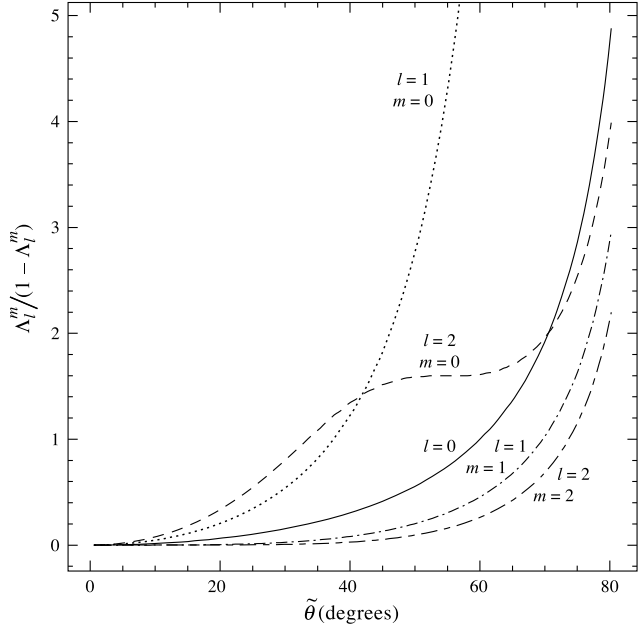}}
\caption{The ratio $\Lambda^m_l(1-\Lambda^m_l)$ of the weighting
  coefficients for various low-degree modes, plotted against the
  angular spot radius $\tilde{\theta}$.}
\end{figure} 

Figure 5 shows the ratio $\Lambda^m_l/(1-\Lambda^m_l)$ determining the
relative contributions to $\nu_{nlm}$ and $\eta_{nlm}$ from the polar
and the equatorial regions for a variety of spot-aligned modes, 
plotted against the
polar angle $\tilde{\theta}$ subtended by the spot radius.  The regions
in which this ratio exceeds $-\eta^{\rm{eq}}_{nl}/\eta^{\rm{p}}_{nl}$,
whose values can be obtained from Figure 4, define the critical spot
sizes above which the corresponding modes are overstable.  Because,
according to Figure 4, the greatest growth rate of the polar component
exceeds the corresponding decay rate of the equatorial component by a
factor of about 4, this simple model predicts that there are
overstable global modes if the spots have angular radii exceeding
about $20^{\circ}$. For $20^{\circ}\lesssim \tilde{\theta} \lesssim
40^{\circ}$, the most unstable modes are axisymmetric quadrupole
modes; if the spots are larger, the most unstable modes are the
axisymmetric dipole modes. Moreover, provided $\tilde{\theta} \lesssim
55^{\circ}$, no nonaxisymmetric mode is unstable.  In this regime
cyclic frequency differences of up to about 20$ \mu$ Hz were computed amongst
dipole modes of like order and degree, and different azimuthal orders.  These
findings are promising, for they augur a more sophisticated theory
that might explain Kurtz's interpretation of the observations.

\begin{figure} \label{fig6}
\centering
\resizebox{0.85\textwidth}{!}{\includegraphics{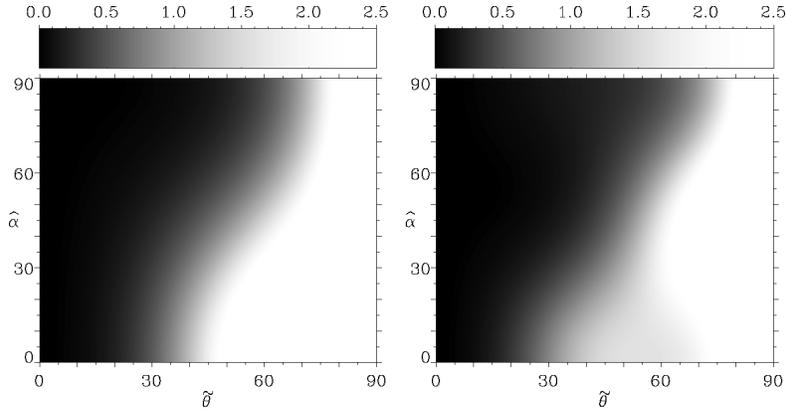}}
\caption{Greyscale plot of the ratio $\Lambda^m_l /(1- \Lambda^m_l)$
  of the weighting coefficients for axisymmetric dipole (left) and
  quadrupole (right) modes inclined by an angle $\hat{\alpha}$ from
  the spot axis. Black denotes stability and white overstability; the
  shade for neutral stability depends on the ratio of the spot-driving
  and equatorial-damping coefficients, as described in the text.}
\end{figure}

When the influence on the oscillation eigenfrequencies of the spots
does not overwhelm that of the star's rotation, the principal axis of
pulsation is not well aligned with the spots.  It has been pointed out by Shibahashi and Takata (1993) and by Bigot and Dziembowski (2002) that this can sometimes be
the case: in particular, the oscillations observed by Kurtz \textit{et
al}., (1992) of HR 3831 may have this property.  In that case
excitation by the spots is less effectual, and one would expect
possibly fewer modes to be excited.  Some of the implications can be
gleened from Figure 6, which comprises grey-scale plots of
$\Lambda^m_l/(1- \Lambda ^m_l)$ against $\tilde{\theta}$ and the angle
$\hat{\alpha}$ subtended by the pulsation axis and the spots (see
Figure 7) for axisymmetric dipole and quadrupole modes, where now
$\Lambda^m_l$ is the first of the integrals in equation (\ref{16}).
Values on the $\tilde{\theta}$ axis correspond to the dotted and
dashed curves in Figure 5.  As $\hat{\alpha}$ increases, the
range of $\tilde{\theta}$ for which quadrupole modes are favoured over
dipole modes shrinks, and disappears at $\hat{\alpha}= 27^{o}$.

\begin{figure} \label{fig7}
\centering
\resizebox{0.45\textwidth}{!}{\includegraphics{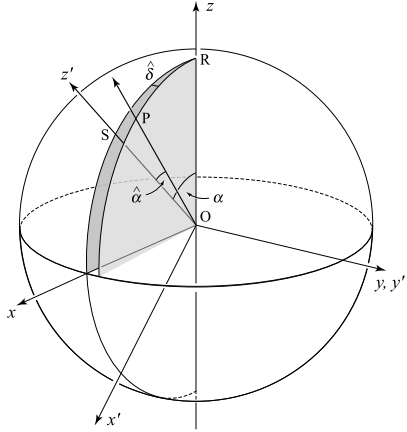}}
\caption{Geometry of a rotating spotted star for a case with
  $\omega_{\rm{s}} > 0$.  The star rotates uniformly about the axis
  OR, and has a pair of antipodal spots about the axis OS, which
  subtends an angle $\alpha$ with OR.  The pulsation axis is OP,
  which, as a result of (small) Coriolis effects, lies (slightly) out of the
  plane of OR and OS.}
\end{figure}

\section{The direct influence of the Lorentz stresses on the
  pulsations}

There have been several studies of the direct influence of the magnetic
fields on modes of stellar oscillation.  Although Lorentz stresses are
very much smaller than gas-pressure gradients throughout almost all
the star, because the field penetrates the surface there must be a
(shallow) region in which the field dominates.  The influence of the
field on the oscillation frequencies is small, but the problem cannot
be treated by standard (nonsingular) perturbation theory.  Instead,
the full dynamical equations must be analysed in the subsurface region
in which the magnitudes of the gas pressure and the Lorentz forces are
comparable.  The problem was considered in the stellar
context by Roberts and Soward (1983), using matched asymptotic
expansions for a plane parallel polytrope -- curvature effects in the
magnetically dominated superficial layers are small -- and
subsequently by Campbell and Papaloizou (1986).  The essentially purely
acoustic waves of the deep interior couple with Alfv\'{e}n waves in
the surface layers, and the Alfv\'{e}n waves propagate downwards into
the star, their wavenumbers increasing with depth as a result of the
increase in density until, it is presumed, they dissipate, never to
return to the surface.  Thus they contribute to the damping of the
modes.  Moreover, the coupling modifies the phase jump that the acoustic
waves experience on (partial) reflection, thereby altering the
frequencies of the resonant modes.

The problem was taken up by Dziembowski and Goode (1996), Bigot
\textit{et al} (2000), Bigot and Dziembowski (2002) and Bigot (2003) 
for more realistic stellar models.   The boundary-layer equations in the
magnetically important surface layers were solved numerically and
patched onto acoustical solutions in the deep interior to obtain a
representation of a complete mode.  The procedure appeared to work
well for kiloGauss magnetic fields, but for the stronger fields that
appear to be present in some stars -- Bagnulo \textit{et al}. (1999)
report 14 kG in HR 3831, although this has been challenged recently by
Kochukhov {\it et al}. (2004) -- the expansions may be invalid.  Saio and
Gautschy (2003, 2004) have extended the expansion to higher order,
which, one hopes, extends the range of usefulness of the results to
higher field strengths.  Frequency perturbations of order 10$ \mu$Hz
for a 1kG magnetic field have been reported by Bigot \textit{et
  al}. (2000) and  Bigot and Dziembowski (2002); Saio and
Gautschy (2004) have reported values twice as great. 

A difficulty with the formal expansions is that they rapidly become
complicated as the expansion is taken to higher order.  To obviate
that difficulty, and at the same time, it was hoped, to increase the
range of magnetic field strengths for which the results are useful,
Cunha and Gough (2000) adopted a much simpler approach: they solved
the full (planar) equations in the magnetic boundary layer
numerically as before, but, instead of expanding the acoustic mode in
the interior about the corresponding solution for a spherically
symmetrical star, they simply adopted the usual asymptotic representation of
that solution expressed as a function of a new radial coordinate
uniformly stretched by a latitudinally (relative to the axis of the
magnetic field) dependent factor chosen to assure phase agreement with
the boundary-layer solution in a common domain of validity of the
inner and outer solutions.  The
amplitude was left unperturbed. The justification for there being hope that the
procedure might be reliable beyond the field strengths at which
Dziembowski and Goode's expansion is useful is that, for the
high-order modes considered, an O(1) phase perturbation modifies the
Eulerian perturbation to the eigenfunction by O(1), which may be
beyond the radius of convergence of the formal expansion, whereas the
vertical wavenumber, on which the frequency principally depends,
changes by only ${\rm{O}}(n^{-1}) \ll1$. However, no serious test of
the procedure was
presented. Moreover, Cunha and Gough carried out the analysis only for
polytropically stratified boundary layers.  Therefore the results
should be treated with some caution. 

An interesting property of the approximate solutions found by Cunha
and Gough (2000) is that there are frequencies near which the
acoustic-Alfv\'{e}n coupling is extremely strong.  Moreover, at these
frequencies the magnetic frequency perturbation, plotted as a function
of mode frequency, experiences a discontinuous drop.  Therefore the
mode frequencies do not adhere to an almost uniformly spaced spectrum,
as one might otherwise have expected. This led Cunha (2001) to propose
the nonuniformity of the coupling as an explanation of the nonuniformity in the spectrum
of HR 1217, an explanation which implied that there should in addition
be another yet undiscovered mode nearer to the location of the
discontinuity but which is too heavily damped to have been seen.  I should
point out that Dziembowski and Goode (1996) and Bigot \textit{et al}.,
(2000) had not found this discontinuous behaviour in their expansions,
which put the phenomenon in doubt (Bigot and Dziembowski, 2002).
However, a new observational search by Kurtz and his collaborators (2002)  
has revealed that the mode is
indeed present, at a lower amplitude than those of the modes
discovered previously, which was encouraging.  Moreover, a subsequent
expansion somewhat similar to that adopted by Dziembowski and his
colleagues but taken to higher order by Saio and Gautschy (2003, 2004) 
did produce such
behaviour, suggesting that Cunha's explanation might at least be
qualitatively correct.          

    The variation of the surface amplitude, $S_{nlm} (\theta, \phi)$, and
the orientation of a mode relative to the axis of rotation of the star
and the line of sight determine the relative amplitudes of the
contribution of that mode to the power spectrum of the observed
oscillations, analysis of which is carried out for the purpose of mode
identification (e.g. Dziembowski and Goode, 1985, 1996; Shibahashi and
Takata, 1993; Takata and Shibahashi, 1995). The perturbation expansions of Dziembowski and Goode
(1996) and Saio and Gautschy (2004), within their range of validity,
determine $S_{nlm}$, which is caused to deviate from the corresponding
spherical-harmonic variation that would have been the structure
had the star been neither spotted nor rotating.  That deviation comes
about partly because the radial variation of the eigenfunction in the
surface layers is modified by the spots in a latitudinally dependent
fashion, and partly because the latitudinally dependent phase jump
experienced by the mode on reflection modifies the interior
eigenfunctions from a spherical-harmonic form.  The approximation
adopted by Cunha and Gough (2000) took account of only the first of these  
effects, an approximation which Montgomery and Gough (2003), 
using an argument based on
an analysis of a rectangular isothermal toy stellar model, subsequently
suggested might lead to substantial inaccuracy, because there might
actually be serious horizontal mode trapping.  Their more recent,
unpublished, work suggests that in a real star wave refraction by the
temperature gradient near the centre mitigates that
trapping; consequently it seems likely that a generalization of the
Cunha-Gough approximation can provide a workable, relatively simple
procedure for computing Ap-star oscillation eigenfunctions.

\section{The orientation of the axis of pulsation}
Mode alignment is induced by the dynamical anisotropy caused by
rotation and the spots, which split the degeneracy   with respect to
azimuthal order $m$ of modes of like order $n$ and degree $l$.  The
associated modifications to the pulsations are probably globally quite small,
although in the surface regions they are certainly significant.  If
spherical symmetry were broken only by rotation, the principal axis of
the modes would be the axis of rotation -- that is to say, with
respect to spherical polar coordinates $(r,\theta, \phi)$ about that
axis the eigenfunctions $\psi_{nlm} (r, \theta, \phi, t)$ could be
written in the approximate form: $\Psi_{ n l} {(r)}~P^m_l (\cos \theta)
{\rm{e}}^{{\rm{i}}(m \phi-\omega t)}$.  The main influence of the
rotation is to assert that alignment and, in addition, to modify the
eigenfrequencies and the eigenfunctions somewhat.  It is normal to
determine the modifications via a degenerate (singular) perturbation
expansion about the nonrotating state.  What follows is a description,
not a derivation, of that theory.

In view of the variational principal that is satisfied by the
eigenfrequencies, the first-order correction to the eigenfrequencies 
can be written in the form $\omega_{\Omega 1 m}=-mC \Omega$, where
$\Omega $ is the angular velocity of the star and $C$ is a factor which is
independent of $m$ and which may be written as an integral depending
on the zero-order eigenfunctions, and the unperturbed density $\rho$ 
(Cowling and Newing, 1948; Ledoux, 1951). It is given by 
$C= \mathcal{I}^{-1} L^{-1} \int\! \left( 2 \xi + L
  \eta \right)\eta \rho r ^2 {\rm{d}}r$,  where $L = l+ \textstyle{1 \over 2}$
and $\mathcal{I} = \int\! \left( \xi^2 + \eta^2 \right) \rho r^2 {\rm{d}}r$
is a measure of the inertia of the mode, $\xi {\rm{~and~}} \eta$ being
respectively the $r$-dependent factors in the vertical and
horizontal displacement eigenfunctions.  For the high-order acoustic
modes of particular relevance to this discussion, $|\eta|/|\xi| = {\rm{O}}
(n^{-1})$, and both $\xi \mbox{ and } \eta$ are oscillatory functions
(with approximately $n$ nodes) that are roughly $\pi/2$ out of phase;
therefore $C={\rm{O}}(n^{-2})$.

The second-order corrections have been discussed by, for example,
Gough and Thompson (1990) and Dziembowski and Goode (1992); they
cannot distinguish between east and west, and therefore they do not
depend on the sign of $m$.  For high-order acoustic modes, at least,
the centrifugal distortion of the star dominates the second-order
correction; consequently the relative second-order correction
$\omega_{\Omega 2}/ \omega_0~$ (of, say, the $m=0$ mode), where 
$\omega_0$ is the eigenfrequency
of the corresponding unperturbed mode, is only weakly dependent on
$n$, and tends to a finite constant as $n \rightarrow \infty$.  In
fact, $\omega_{\Omega 2} \sim n \pi \left(\Omega^2/ \sigma^2 \right) D
\sigma$, where $ \sigma^2 = GM/R^3$ is the characteristic global dynamical
frequency of the star and $D$ is a dimensionless $m$-dependent integral 
of order unity, which depends on (predominantly) the zero-order 
oscillation eigenfunctions.

It is convenient to separate the frequency perturbation into its mean
(over $m$) value $\overline{\omega}_{\Omega 2}$ and an $m$-dependent
residual $\omega_{\Omega 2 m}$ satisfying $\sum_m \omega_{\Omega2m}
=0$.  The mean value arises principally from the effect of the
centrifugal force on the spherically averaged structure of the star,
and, in the present state of the subject, cannot be distinguished from
uncertainties in the structure of the corresponding nonrotating star,
and so need not concern us here. If one specializes to dipole modes,
which are the most pertinent to roAp stars, one can write
$\omega_{\Omega 2 0}=\omega_{\Omega 2}$, and $\omega_{\Omega 2 m}= - {1
  \over 2} \omega_{\Omega 2}$ when $m= \pm 1$

Roughly speaking, the differential equation describing the oscillation
modes $(\Psi_{k}, \omega_{k})$ can be written in the form

\begin{equation} \label{21}
\mathcal{L} \Psi_{k} + \mathcal{R} \Psi_{k} - \lambda_{k}
\Psi_{k} =0~,
\end{equation}
where $\Psi_{k}$ represents eigenfunction $k$ of the $(2l+1)$-fold
multiplet of modes of given $n$ and $l$, and $\lambda_{k} =
\omega^2_{k}$ is the square of its eigenfrequency; $\mathcal{L}$ is
the differential operator describing the oscillations
of the corresponding nonrotating star and $\mathcal{R}$ is the
operator describing the effect of rotation. For the purposes of this
part of the discussion it is adequate to adopt the adiabatic
approximation, in which case $\mathcal{L}$ is self-adjoint.
(Generalization to nonadiabatic pulsations is straightforward, but
because one must then deal with a non-self-adjoint operator the
argument is a little more cumbersome.)  To leading order in the
rotational perturbation, eigenfunctions $\Psi_{k}$ can be expressed
as a linear combination of the (orthogonal) zero-order eigenfunctions
$\Psi_{nlm}^{(0)}$:

\begin{equation} \label{22}
\Psi_{k}=\sum_ma_{km}\Psi^{(0)}_{nlm}~,~\mathcal{L}\Psi^{(0)}_{nlm}
-\lambda^{(0)}_{k} \Psi^{(0)}_{nlm}=0~,
\end{equation}
which are normalized such that
$\langle\Psi^{(0)*}_{nlm^\prime}\Psi^{(0)}_{nlm}\rangle=\delta_{m^\prime
m}$, where the angular brackets denote weighted integration over the
star with the weight function with respect to which $\mathcal{L}$ is self-adjoint, 
and the asterisk denotes complex conjugate. Substituting equations
(\ref{22}) into equation (\ref{21}), writing
$\lambda^{(1)}_{k}=\lambda_{k} - \lambda^{(0)}_{k}$,
premultiplyng by $\Psi^{(0)}_{nlm^\prime}$ and integrating yields

\begin{equation} \label{23}
\mathcal{R}{\textbf{\textit a}}_{k} - \lambda^{(1)}_{k}{\textbf{\textit a}}_{k} = {\bf{0}}~,
\end{equation}
where $\mathcal{R}$ now represents the matrix with components
$\mathcal{R}_{m^\prime m}=\langle
\Psi^{(0)}_{nlm^\prime}\mathcal{R}\Psi^{(0)}_{nlm}\rangle$ and
${\textbf{\textit a}}_{k}$ is the $(2l+1)$-dimensional vector with components
$a_{km}$. In view of the discussion in the previous paragraph, $\mathcal{R}$ is diagonal, and in the simple case of
dipole modes,

\begin{equation} \label{24}
\mathcal{R}=-c\Omega {\rm A}+\omega_{\Omega 2}{\rm B}~,
\end{equation}
where

\begin{equation} \label{25}
{\rm{A}}=\left(
\begin{array}{ccc}
-1&0&0\\
0&0&0\\
0&0&1
\end{array}
\right)
,~~{\rm{B}}=
\left(
\begin{array}{ccc}
-\frac{1}{2}&0&0\\
0&1&0\\
0&0&-\frac{1}{2}
\end{array}
\right)~.
\end{equation}

\vspace{5mm}
The total effect of rotation on a group of modes of oscillation of
like $n$ and $l$ can be summarized thus: The geometrical structure of
each of the modes is perturbed only slightly, but the orientations of
the modes are constrained relative to the axis of rotation. The
frequencies are perturbed by centrifugal distortion by an amount that
depends on the magnitude of $m$, but centrifugal Coriolis effects are
proportional to $m$, and two modes of like $|m|$ (with the same
amplitude), having structure proportional to $P^m_l \left( \cos \theta
\right) \exp \left[ -{\rm{i}} \left( \omega_0 + \omega _{\Omega 2}
\right) t - {\rm{i}} m \left(\phi+ C\Omega t \right)\right]$ and
$P^m_l \left(\cos \theta\right) \exp \left[ -{\rm{i}}\left( \omega _0+
\omega _{\Omega2} \right) t+ {\rm{i}}m \left( \phi+ C \Omega t \right)
\right]$ can be combined to yield \/ two \/ essentially \/ identical
standing \/ modes \/ having \/ structure \/ proportional \/ to
\newline 
$P^m_l \left(\cos \theta\right) \cos \left[ \left( \omega_0
    +\omega_{\Omega 2}\right) t \right] \!\! \phantom{1}^{\sin}_{\cos} \left[m
 \left( \phi+ C \Omega t \right) \right]$,
which, as a result of the Coriolis force, precess about the axis of
rotation with angular velocity $-C\Omega$.

The effect of the asphericity associated with the spots is formally
similar to that produced by the centrifugal force.  If the star were
not rotating, the modes would be aligned with the spots, and could not
distinguish between east and west.  Therefore, with respect to
spherical polar co-ordinates $(r, \theta^{\prime}, \phi^{\prime})$
about the spot axis, the eigenfunctions are given approximately by
$\Psi_{nlm'} ( r, \theta^{'}, \phi^{'},t ) = \psi_{nl} (r) P ^{m'}_l (
\cos \theta^{'} ) \exp[{\rm{i}(m^{'}\phi^{'}-\omega t)]}$, where
$\omega = \omega_0 + \overline{\omega}_{{\rm{s}}}+\omega_{{\rm{s}}m}$
and $\sum_m \omega_{{\rm{s}}m} =0$.  For dipole modes,
$\omega_{{\rm{s}}0}=\omega_{\rm{s}}$ and $\omega_{{\rm{s}m}}= - {1
\over 2} \omega_{{\rm{s}}}$. By analogy with the centrifugal effects,
if $\mathcal{M}^{\prime}$ is the matrix corresponding to $\mathcal{R}$ associated with the spot,
but referred to the axis of the spots, then $\mathcal{M}^\prime$ is diagonal,
and is given by $\mathcal{M}^\prime=\omega_{\rm s}{\rm B}$.

When the effects of both the spots and the rotation are taken into
account, the dipole eigenfunction can again be expressed to leading
order as an axisymmetric dipole $\cos \theta^{\prime\prime}$ with
respect to polar coordinates
$(\theta^{\prime\prime},\phi^{\prime\prime})$ about an axis whose
orientation is determined by the relative strengths of the
perturbations. If Coriolis effects were ignored, then, by symmetry,
that axis would lie in the plane of the axis of rotation and the axis
of the spots, inclined, say, by an angle $\hat{\alpha}$ from the axis
of the spots. I call this axis ${\rm{O}\hat{\rm{P}}}$. Then if
${\hat{\rm{R}}}$
represents the matrix that rotates $\mathcal{M}{'}$ by $\hat{\alpha}$
from the axis of the spots to the pulsation axis, and ${\rm R}{'}$ the matrix
that rotates $\mathcal{R}$ from the rotation axis to the pulsation
axis, the matrix ${\rm{R}}^{\prime{\rm t}}\mathcal{R} {\rm{R}}^{'} +
{\hat{\rm{R}}}^{\rm t}\mathcal{M}^{\prime} {\hat{\rm
    R}}-\lambda^{(1)}_{k}{\rm I}$, where ${\rm I}$ is
the unit matrix and the superscript t represents transpose, is
diagonal. The axis of pulsation is most easily found as the direction
of the appropriate eigenvector of $\mathcal{R}+\mathcal{M}$, referred to the
rotation axis, or $\mathcal{R}^{\prime}+\mathcal{M}^{\prime}$ referred
to the spot axis. When the (small) Coriolis effects are included, the
pulsation axis OP deviates slightly from ${\rm{O}\hat{\rm{P}}}$
(Figure 7). There is an axisymmetric dipole mode
aligned with that axis, and two perpendicular linearly independent
dipole oscillations precessing about it.  Further
details are given by Bigot and Dziembowski (2002).

It has usually been assumed that the spot perturbation is much greater
in magnitude than the centrifugal perturbation: $|\omega_{\rm{s}}| \gg
| \omega_{\Omega 2}|$. In that extreme it is evident that rotation has
  only a small influence on the dynamics, and there is always a mode
  that is almost aligned with the spots.  That is essentially the
  oblique-rotator model discussed by Kurtz (1982), and is more-or-less what
  I had in mind when discussing the mode excitation in \S4 and \S5.
  Viewed in the frame of the rotating star, an axisymmetric mode of
  degree $l$ and frequency $\omega$ in the frame of the rotating star can be represented as a sum of oscillations of the form 
$\sum_m A_m P_l^m (\cos \theta){\rm{e}}^{{\rm{i}}(m \phi-\omega t)}$
about the axis of rotation, in which the (normalized) coefficients
$A_m$ depend solely on $m$ (and $l$), and the inclination $\alpha$ of the
spot axis from the rotation axis.  Viewed from the earth,
the different contributions to the sum have different frequencies, approximately
equal to $\omega + m \Omega$, and can therefore be distinguished in a
power spectrum of the observed oscillation signal. Their observed relative
amplitudes ${\it a}_m/{\it a}_0$, which depend on $A_m$ and the
inclination $\beta$ of the rotation axis relative to the line of sight, constrain $\alpha$ and $\beta$, and
provide a 
consistency check of the oblique-rotator description (Kurtz
and Shibahashi, 1986): if the description is correct,
then ${\it a}_{-1} = {\it a}_{+1}$.

It seems however, that not all roAp stars are that simple. In particular, it appears that in HR 3831, $a_{-1}\neq a_{+1}$ (Kurtz
{\textit{et al}}., 1997). Shibahashi and Takata (1993) and  Bigot and Dziembowski (2002) have attributed that inequality
to the influence of the Coriolis effects.  They pointed out that if
the star is rotating rapidly enough the spot-induced perturbations do
not overwhelm the rotational perturbations, and the axis of pulsation
deviates substantially from the spot axis.  Then Coriolis effects
cause the amplitudes of the signal associated with the $m= -1$ and the
$m=+1$ components to differ. 
Moreover, there are additional components in the spectrum arising from
the distortion of the eigenfunctions away from their zero-order
spherically harmonic state (considered, for example, by Shibahashi and
Takata, 1993, and Takata and Shibahashi, 1995, although they did not
take centrifugal effects into account), the details of which I shall
not discuss here.

When centrifugal effects are significant, the axis of the principal
(nonprecessing) mode is no longer well aligned with the spots, and
therefore the $\kappa$ mechanism is less effectual in exciting the
oscillations (see Figure 6).  Perhaps this is why roAp stars are
generally observed to be slow rotators.  In the extreme case when 
$|\omega_{\rm{s}} | \ll | \omega _{\Omega 2}|$ and, as in Figure 2, $\alpha \simeq \pi/2$, the principal mode is
almost aligned with the rotation axis.  But there are also two other linearly
independent dipole modes, which precess about that axis,  and
whose axes of symmetry therefore pass through the spots.  This is the case
that was considered by Dolez and Gough(1982), who suggested that if the modes
are indeed excited in the spots (Dolez and Gough did not
actually find net excitation, principally because of a flaw in the
opacity tables available at that time), when they are aligned, then, because the
growth time when they are aligned and the decay time when they are not
aligned are both likely to be much shorter than the time it takes for the
mode to precess across the spot, the modes will always be found
approximately aligned with the spots and might thereby be presumed not to be precessing. That possibility was
dismissed by Dziembowski and Goode (1985) as being unnecessarily
complicated, but maybe in the future it will be found that
in some stars such complication is present.

\section{Conclusion}
It is my opinion that a theory of the roAp-star phenomenon is
beginning to emerge.  The stars are slow rotators.  They are spotted as
a result of there being a large-scale predominantly dipole magnetic
field pervading the star, inclined from the axis of rotation, which
suppresses subsurface convection to form two antipodal spots where
the field is almost vertical.  This permits the establishment of
chemical abundance anomalies brought about a combination of radiative
levitation and gravitational settling against diffusion.  There might
also be a stellar wind from the spot, where the magnetic field lines
are effectively open. This would modify the chemical abundances, as also
might transport by fingering in regions where the mean molecular mass
increases upwards, although this process, unfortunately, is not well understood.
The oscillations are low-degree $\rm{p}$ models of high order, which
are probably axisymmetric with axes more-or-less aligned with the spots.
They are excited by the $\kappa$ mechanism in the spots, and damped by
convection elsewhere.  Only a few high-order modes are overstable, in
accord with observation. Some calculations have found overstability also in some
 relatively-low-order modes, with much lower growth rates; perhaps
 such oscillations are present too, but at amplitudes too low yet to
 have been observed.

An important matter that has never been seriously addressed is an
explanation of the amplitudes of the modes.  If the modes are
overstable, their growth must be limited by some nonlinear process.
Is that process similar to that which limits the growth of the
oscillations of Cepheids and RR Lyrae stars?  If so, why does it
operate effectively at so low and amplitude?

Only when the spots are sufficiently large, and the rotation of the
star is sufficiently small, can axisymmetric oscillation modes be
dynamically locked with their axes close enough to the spots for excitation to
dominate over damping, causing overstability; nonaxisymmetric modes
precess, due to Coriolis effects, and cannot be firmly locked.
Because the oscillation amplitude is greatest on the axis of symmetry,
overstability is most likely to result when the modes are aligned with
the spots. Of course, the larger the spots, the less precise need be
the alignment, and in the fullness of time a robust theoretical
criterion for overstability might be found: a relationship between spot size and  inclination of the oscillation axis
from the spots.  Nonaxisymmetric
modes have lower growth rates, and are overstable only if the spots
are very large; that they appear not to be observed would set an upper
limit to the size of the spots, once a robust theory is available.

In some circumstances the stellar rotation might be so large that the
principal axis of pulsation is more-or-less aligned with the rotation
axis, and not with the spots; in that case, it might be that other
modes of the same degree precess slowly across the spot and become
overstable only when they are appropriately orientated, giving the
impression of dynamical alignment.

The theoretical models are at present extremely primitive. The
influence of the spots on the pulsations has been accommodated in a
piecemeal way, and must in future be properly incorporated into the
pulsation dynamics to produce a plausible quantitative model.  There
has been substantial progress in studying the influence of the Lorentz
forces on the pulsations, but only in stellar models without
spots. Lateral inhomogeneity of the background state,
convection-pulsation interactions, Lorentz forces and nonadiabaticity
must all be combined in a consistent way.  And our understanding of
the dynamics of the oscillations in the atmosphere, where nonlinearity
in the dynamics can be important, and where the processes of amplitude limitation are probably the most effectual, must be improved.  Only then can one
reliably use future observations of mode frequencies and amplitudes, intensity-velocity phase relations and
nonlinearities in the light-curves to draw reliable inferences about
the structure of the stars.  We have many (but not all) of the ingredients in hand.
Therefore a preliminary theory might be almost in sight.

\section{Acknowledgements}
I am very grateful to J. Gilbert and D. Sword for preparing the
typescript, and to R. Sword and G. Houdek for their help in preparing
the diagrams.

\section{references}
 
\noindent Bagnulo, S., Landolfi, M. and Landi Degl'Innocenti, M. 1999, Astron. 

Astrophys., {\bf 343}, 865--871

\noindent Balmforth, N.J., Cunha, M.S., Dolez, N., Gough, D.O.  and Vauclair, S. 2001, 

Mon. Not. R. Astron. Soc., {\bf 323}, 362--372

\noindent Bernstein, I. B., Frieman, E. A., Kruskal, M. D. and Kulsrud, R. M. 1958, Proc. 

Roy. Soc., {\bf	A 244}, 17--40

\noindent Bigot, L.  2003, {\it International Conference on magnetic fields in O, B and A stars}, 

(ed.  L.A. Balona, H.F. Henrichs \& T. Medupe), {    Astron. Soc. Pacific  

Conf. Ser.},  {\bf 305}, 73--82

\noindent Bigot, L. and Dziembowski, W.A. 2002, Astron. Astrophys., {\bf 391}, 235--245

\noindent Bigot, L., Provost, J., Berthomieu, G., Dziembowski, W.A. and Goode, P.R. 

2000,   { Astron. Astrophys.},  {\bf 356},
218--233

\noindent Blanchflower, S. and Weiss, N. 2002,  Phys. Lett. A, {\bf 294}, 297-303

\noindent Campbell, C. G. and Papaloizou, J. C. B.  1986,  Mon. Not. R. Astron. Soc.,

{\bf 220}, 577-591

\noindent Chandrasekhar, S. 1961, {\it Hydrodynamic and Hydromagnetic Stability} (Oxford 

Univ. Press)   

\noindent Cowling, T.G. and Newing, R.A. 1949, Astrophys. J., {\bf 109}, 149--158

\noindent Cunha, M. S.  1998,  Contrib. Astron. Obs. Skalnate Pleso, {\bf 27}, no 3, 272-279

\noindent Cunha, M.S. 2001, { Mon. Not. R. Astron. Soc.}, {\bf 325}, 373--378

\noindent Cunha, M.S. and Gough, D.O. 2000, Mon. Not. R. Astron. Soc., {\bf 319}, 1020--

1038

\noindent Dolez, N. and Gough, D.O. 1982, {\it Pulsations in classical and cataclysmic 

variable stars} (ed. J.P. Cox \&C.J. Hansen, JILA, Boulder), pp 248--256
 
\noindent Dziembowski, W.A. and Goode, P.R. 1985, {Astrophys. J.}, {\bf  296}, L27--L30

\noindent Dziembowski, W.A. and Goode, P.R. 1992 Astrophys. J., {\bf 394}, 670--687

\noindent Dziembowski, W.A. and Goode, P.R. 1996, Astrophys. J., {\bf 458}, 338--346

\noindent Gautschy, A. and Saio, H.  1998, Mon. Not. R. Astron. Soc., {\bf 301}, 31--41

\noindent Gough, D.O. 2002 {\it Radial and Nonradial Pulsations as Probes of Stellar Physics}, 

({ Proc. IAU Colloq.}, {\bf 185}, ed. C. Aerts, T.R. Bedding \& J. Christensen-

Dalsgaard), { Astron. Soc. Pacific Conf. Ser.},  {\bf 259}, 37--57 

\noindent Gough, D. O. and Tayler, R. J. 1966, Mon. Not. R. Astron. Soc. {\bf 133}, 85 

\noindent Gough, D.O. and Thompson, M.J. 1990, Mon. Not. R. Astron. Soc. {\bf 242}, 

25--55 

\noindent Kochukhov, O., Drake, N. A., Piskunov, N. and de la Reza, R. 2004, Astron. 

Astrophys., {\bf 424}, 935-950 

\noindent Kurtz, D. W. 1982, Mon. Not. R. Astron. Soc., {\bf 200},  807-859

\noindent Kurtz, D. W.  1990, Ann. Rev. Astron. Astrophys., {\bf 28} (A91-28201 10-90)

\noindent Kurtz, D. W. and Shibahashi, H.  1986,  Mon. Not. R. Astron. Soc., {\bf 223}, 557

\noindent Kurtz, D. W., Kanaan, A., Martinez, P. and Tripe, P. 1992, Mon. Not. R. 

Astron. Soc., {\bf 255}, 289

\noindent Kurtz, D. W.  et al. 1997, Mon. Not. R. Astron. Soc., {\bf 287},  69--78

\noindent Kurtz, D. W. et al. 2002, Mon. Not. R. Astron. Soc., {\bf 330}, L57-L61

\noindent Lebovitz, N. R. 1965, Astrophys. J., {\bf 142}, 229-242

\noindent Lebovitz, N. R. 1966, Astrophys. J., {\bf 146}, 946-949 

\noindent Ledoux, P. 1951, Astrophys. J., {\bf 114},  373--384

\noindent Martinez, P. and  Kurtz, D. W. 1995, {\it Astrophysical applications of stellar 

pulsation} (ed R. S. Stobie \& P.A. Whitelock) Proc. IAU Colloq.  {\bf 155},  

Astron. Soc. Pacific Conf. Ser., {\bf83}, 58

\noindent Montgomery, H.M. and Gough, D.O. 2003, {\it Asteroseismology across the 

HR diagram}, (ed. M.J. Thompson, M.S. Cunha \& M.J.P.F.G. Monteiro, 

Kluwer, Dordrecht),   pp 545--548 

\noindent Newcomb, W.A. 1961, Phys. Fluids, {\bf 4}, 391--396

\noindent Roberts, P. H. and Soward, A. M. 1983, Mon. Not. R. Astron. Soc., {\bf 205}, 

1171-1189

\noindent Saio, H. and  Gautschy, A.  2003,   {\it International Conference on magnetic fields in 

O, B and A stars}, (ed.  L.A. Balona, H.F. Henrichs \& T. Medupe ), 

{   Astron. Soc. Pacific  Conf. Ser.}, {\bf 305}, 140--145

\noindent Saio, H. and  Gautschy, A.  2004, Mon. Not. R. Astron. Soc., {\bf 350}, 485-505

\noindent Schwarzschild, K. 1906, Nach. K\"on. Gesellsch. Wiss. G\"ottingen, {\bf 195}, 41--53

\noindent Shibahashi, H. and Takata, M. 1993, Publ. Astron. Soc. Japan, {\bf 45}, 617--641

\noindent Takata, M. and Shibahashi, H. 1995, Publ. Astron. Soc. Japan, {\bf 47}, 219--231

\noindent Tayler, R.J. 1961, J. Nucl. Energy, {\bf 3}, 266--272

\noindent Vauclair, S., Dolez, N. and Gough, D.O. 1991  {Astron. Astrophys.}, {\bf 252}, 618--624

\end{document}